\documentclass{ws-procs975x65}
\usepackage{enumitem}
\usepackage{wrapfig}

\def\beq{\begin{equation}}
\def\eeq{\end{equation}}

\begin{document}

\title{Observers' measurements of time and length in premetric electrodynamics}

\author{Christian Pfeifer} 

\address{Laboratory of Theoretical Physics, Institute of Physics, University of Tartu, W. Ostwaldi 1\\
50411 Tartu, Estonia\\
E-mail: christian.pfeifer@ut.ee}

\begin{abstract}
The notion of observers' and their measurements is closely tied to the Lorentzian metric geometry of spacetime, which in turn has its roots in the symmetries of Maxwell's theory of electrodynamics. Modifying either the one, the other or both ingredients to our modern understanding of physics, requires also a reformulation of the observer model used. In this presentation we will consider a generalized theory of electrodynamics, so called local and linear premetric, or area metric, electrodynamics and its corresponding spacetime structure. On this basis we will describe an observer's measurement of time and spatial length. A general algorithm how to determine observer measurements will be outlined and explicitly applied to a first order premetric perturbation of Maxwell electrodynamics. The latter contains for example the photon sector of the minimal standard model extension. Having understood an observer’s measurement of time and length we will derive the relativistic observables time dilation and length contraction. In the future a modern relativistic description of the classical test of special relativity shall be performed, including a consistent observer model.
\end{abstract}


\bodymatter

\section{From electrodynamics to relativity}
The origins of special and general relativity and, in particular, of the relation between observers by Lorentz transformations, lies in Maxwell's theory of electrodynamics. It predicts the propagation of light on the integral curves of the directions forming the null cones cones of a Lorentzian metric, and the Lorentz transformations are those transformations which leave this structure invariant, globally for the flat Minkowski metric, locally for any non-flat Lorentzian metric g encoding gravity. Thus historically, first there was a viable relativistic matter field theory which described the behaviour of the electromagnetic field correctly, and then, a theory of gravity consistent with the relativity principles of the matter field theory was constructed \cite{ThePrincipalOfRelativity}.

Modifications and extensions of general relativity, and of the field theories forming the standard model of particle physics, are numerous and disperse in various directions. However, it is rarely discussed what happens to the description of observers and their measurements, when one changes the theoretical models of gravity or matter. In approaches where gravity is still encoded into a spacetime metric and matter fields are minimally coupled to this metric, such as f(R)-gravity or other modifications of the Einstein-Hilbert action only involving the metric \cite{Nojiri:2017ncd}, the observer description from general relativity is still applicable. But, when one changes the field which encodes gravity away from the metric, or introduces couplings of this field to matter fields, like in the standard model extension (SME) framework \cite{Colladay:1998fq,Bluhm:2005uj}, the Robertson-Mansouri-Sexl (RMS) framework \cite{Robertson:1949zz,Mansouri:1977zz}, Finsler geometries \cite{Pfeifer:2011xi} or premetric electrodynamics \cite{Hehl,Schuller:2009hn}, the influence of the added concepts on observers, their measurements and their relation between each other has to be investigated and a consistent observer model has to be constructed.

Here we summarize how one obtains an observer's measurement of time and spatial lengths starting from a theory of electrodynamics in the following algorithm, using the techniques developed in the articles \cite{Raetzel:2010je,Pfeifer:2014yua}: 
\begin{enumerate}[label=\Roman*.]
	\item Derive the geometric optics limit of the theory of electrodynamics.
	\item Derive the Lagrange functions $L^\#$ and $L^*$, which define the motion of massless and massive particles on the manifold, from the geometric optic limit.
	\item Use $L^*$ to realize the clock postulate, i.e.\ to identify the proper time normalization of observer worldlines~$x$ by choosing $L^*(\dot{x})=1$.
	\item Model the radar experiment by demanding that for a spatial direction $X$ the vectors $N^+ = \ell_{\dot{x}}(X)\dot{x} + X$  and $N^- = \tilde \ell_{\dot{x}}(X)\dot{x} + X$ are the tangents of the light rays of the radar signal, i.e.\ are null-vectors of $L^\#$.
	\item A solution of $L^\#(N^\pm)=0$ defines the radar length $\mathcal{L}_U(X) = \ell_{\dot{x}}(X) + \tilde \ell_{\dot{x}}(X)$ an observer on worldline $x$ associates to an object represented by the vector~$X$.
\end{enumerate}
To demonstrate how the general algorithm works in practice we apply it to first order modification of Maxwell electrodynamics caledl weak premetric electrodynamics.

Throughout this article we use the convention that $\eta = \mathrm{diag}(1,-1,-1,-1)$.
\section{Weak premetric electrodynamics}
The field equations of weak premetric electrodynamics are \cite{Gurlebeck:2018nme}
\begin{align}\label{eq:edynsme}
(\eta^{ac}\eta^{bd} + \mathcal{K}^{abcd})\partial_bF_{cd}=0, \quad F_{cd} = \partial_{[c}A_{d]}\,,
\end{align}
the tensor $\mathcal{K}^{abcd}$ parametrizes the deviations from metric electrodynamics and has the following properties $\mathcal{K}^{abcd}=\mathcal{K}^{[ab][cd]}=\mathcal{K}^{[cd][ab]}$ and $\mathcal{K}^{a[bcd]}=0$.

\textbf{Step I. Plane waves: The geometric optic limit.} Wave covectors $k$ of the plane waves solving the field equations are the roots of the Fresnel polynomial~$\mathcal{G}(k)$ \cite{Rubilar:2007qm}
\begin{align}\label{eq:FresnelPert}
\mathcal{G}(k) = \eta^{-1}(k,k)^2 - \eta^{-1}(k,k)\mathcal{K}(k,k) + \frac{1}{2}\big(\mathcal{K}(k,k)^2 -\mathcal{J}(k,k,k,k) \big)\,.
\end{align}
The tensor fields appearing are derived from the perturbation tensor defining the field equations
\begin{align}
	\mathcal{K}(k,k) &= \mathcal{K}^{ac}k_ak_c = \mathcal{K}^{abc}{}_{b}k_ak_c,\\
	\mathcal{J}(k,k,k,k) &= \mathcal{J}^{acef}k_ak_ck_ek_f = \mathcal{K}^{abcd}\mathcal{K}^{e}{}_{b}{}^f{}_d k_a k_c k_ek_f\,.
\end{align}
Observe that, in order to study the correction in the propagation of light predicted by the perturbation of the field equations to first order, it is necessary to consider the Fresnel polynomial to second order. For the following, in particular the condition $\det(\partial_{k_a}\partial_{k_b}\mathcal{G}(k))\neq 0$ must hold \cite{Raetzel:2010je}.

\textbf{Step II.: The determination of the Lagrange functions $L^\#$ and~$L^*$.} The Lagrangian which determines the propagation of light is given by the dual polynomial of $\mathcal{G}(k)$, i.e.\ by the unique polynomial satisfying $L(\dot x(k)) = Q(k)\mathcal{G}(k)$, where $\dot x^a(k) = \partial_{k_a}\mathcal{G}(k)$. Employing $\dot x^\flat = \eta(x,\cdot)$ we find
\begin{align}
	L^\#(\dot x) 
	&= \eta(\dot x,\dot x)^2 + 3 \mathcal{K}(\dot x^\flat,\dot x^\flat) \eta(\dot x,\dot x) \nonumber\\
	&+ \frac{1}{2}(3 \mathcal{J}(\dot x^\flat,\dot x^\flat,\dot x^\flat,\dot x^\flat) + 4 \mathcal{K}(\dot x^\flat,\dot x^\flat)^2 + 5 \mathcal{K}_{ac} \mathcal{K}^{c}{}_{b} \dot x^{a} \dot x^{b} \eta(\dot x,\dot x))\label{eq:Lsharp}\\
	&=\eta(\dot x,\dot x)^2 + h_1(\dot x,\dot x) \eta(\dot x,\dot x) + h_2(\dot x,\dot x,\dot x,\dot x)\,.\label{eq:Lsharp2}
\end{align}
The abbreviations $h_1$ and $h_2$ were introduced to display the radar length formula later in a compact form.
The motion of observers and their geometric clock is given by a second Lagrangian $L^*$ which can be obtained from the Helmholtz action 
\begin{align}
S[x,k,\lambda] = \int \mathrm{d}\tau\ \Big( k_a\dot x^a - \lambda \ln\big(\mathcal{G}(\tfrac{k}{m})\big)  \Big),
\end{align}
by eliminating $k$ and the Lagrange multiplier $\lambda$ with help of the corresponding equations of motion. Doing so we find the length functional
\begin{align}\label{eq:Lstar}
	S[x] = \int \mathrm{d}\tau\ L^*(\dot x) = \int \mathrm{d}\tau\ \sqrt{\eta(\dot x,\dot x)}\bigg(1+\frac{1}{4}\frac{\mathcal{K}(\dot x^\flat, \dot x^\flat) }{\eta(\dot x, \dot x)}\bigg)\,.
\end{align}
Physically this parametrization invariant action integral measures the observer's proper time, and mathematically speaking $L^*$ is a Finsler function.

\textbf{Step III.: The clock postulate.} An observer worldline is a cuvre $x(\tau)$ whose tangent satisfies $L^*(\dot x) = 1$, or explicitly $\eta(\dot x, \dot x) = 1 - \frac{1}{2}\mathcal{K}(\dot x^\flat,\dot x^\flat)$. The directions $X$ which are spatial w.r.t. $x$ satisfy $X^a \partial_{\dot x^a}L^* = \eta(\dot x, X) + \frac{1}{2}\mathcal{K}(\dot x^\flat, X^\flat) = 0$.

\section{The radar length}

\begin{wrapfigure}{r}{0.3\textwidth}
	\vspace{-20pt}
	\begin{center}
	\includegraphics[width=0.2\textwidth]{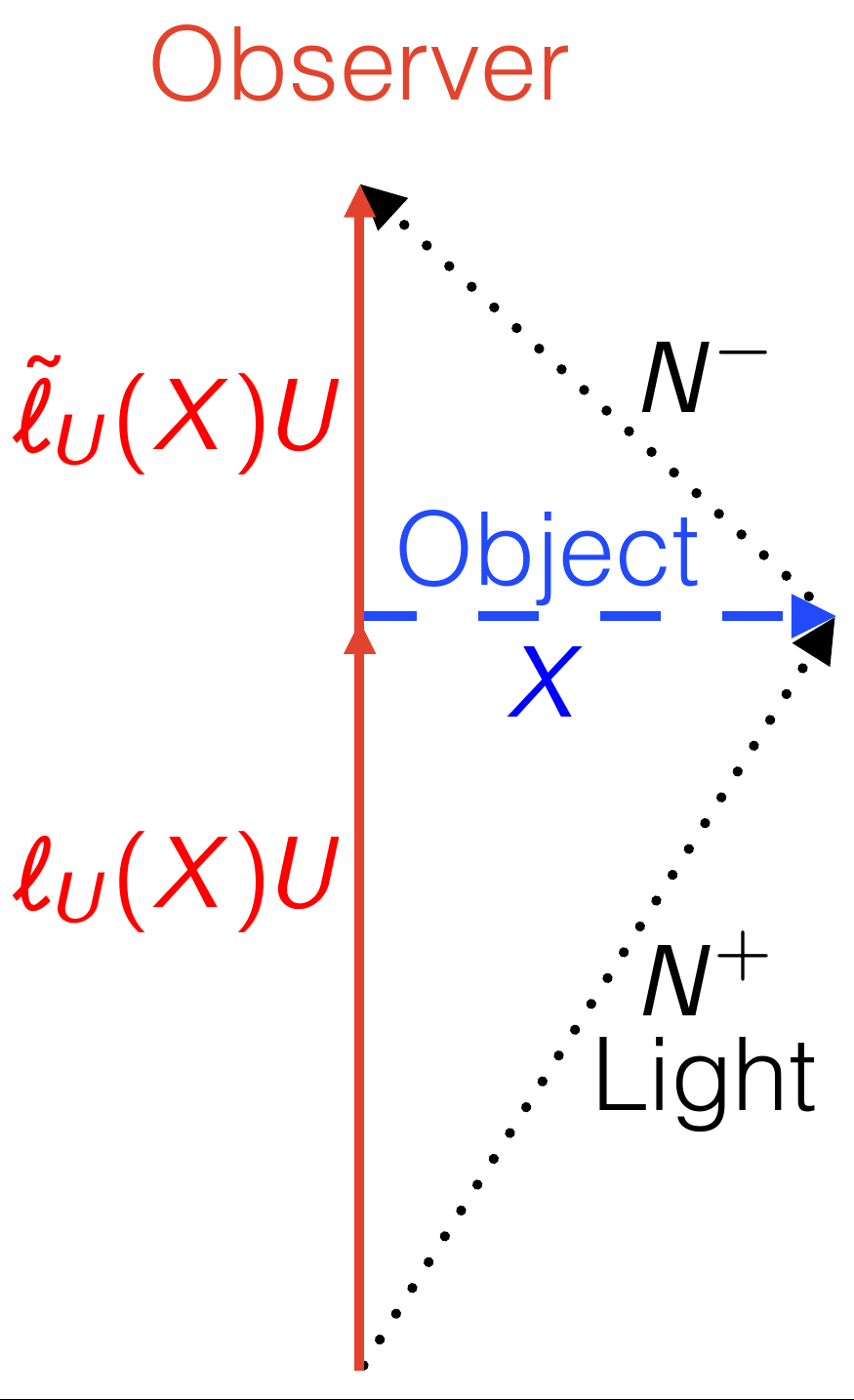}
	\end{center}
	\vspace{-15pt}
	\caption{A radar experiment}
	\label{fig:1}
	\vspace{-12pt}
\end{wrapfigure}
\textbf{Step IV.: Setup the radar experiment.} An observer on a wordline $x(\tau)$ with tangent $\dot x$ emits light along the curve with tangent $N^+$ towards the end of an object, which is modelled by a vector $X$ spatial w.r.t. $\dot x$. There the light gets reflected and propagates back to the observer along the curve with tangent $N^-$. The observer measures the time of flight of the light between its emission and return. This time is the radar length the observer associates to the object. The tangents of the light curves $N^+ = \ell_{\dot{x}}(X)\dot{x} + X$ and $N^- = \tilde{\ell}_{\dot{x}}(X)\dot{x} - X$ must satisfy $L^\#(N^+) = 0 = L^\#(N^+)$, which can be solved for $\ell_{\dot{x}}(X)$ and $\tilde {\ell}_{\dot{x}}(X)$. 

\textbf{Step V.: Solve for the radar length.} The radar length then is given by $\mathcal{L}_{\dot x}(X) = \ell_{\dot{x}}(X) + \tilde \ell_{\dot{x}}(X)$.

The influence of $L^\#$ on the radar experiment is obvious, however it is important to stress that also $L^*$ enters the calculation by the demands that $\dot x$ satisfies $L^*(\dot x) = 1$ and the fact that $X$ shall be spatial w.r.t. $\dot x$, see the comments below \eqref{eq:Lstar}.

Using the explicit form of $L^\#$ and $L^*$ in \eqref{eq:Lsharp} and \eqref{eq:Lstar}, the desired radar length is given by, see \cite{Gurlebeck:2018nme} for the derivation,
\begin{align}
	\mathcal{L}_{\dot x}(X)_{\sigma\tilde{\sigma}}
	&= 2 \sqrt{- \eta(X,X)}\\
	&+ \frac{\epsilon}{4} \frac{\ \bigg(\ \sigma \sqrt{B(X,\dot x)}  + \tilde \sigma \sqrt{ \tilde B(X,\dot x)}\ \bigg)}{\sqrt{- \eta(X,X)}} 
	- \frac{\epsilon}{2} \frac{h_1(X,X) - 2 h_1(\dot x,\dot x)\eta(X,X)}{\sqrt{ - \eta(X,X)}}\nonumber\,.
\end{align}
The abbreviations $B$ and $\tilde B$ are lengthy functions of the perturbation tensors $h_1$ and $h_2$ defined in \eqref{eq:Lsharp2}, they are displayed in \cite{Gurlebeck:2018nme}. The labels $\sigma$ and $\tilde \sigma$ can each take the values $+1$ or $-1$ and label which polarization of light has been used for $N^+$ resp. $N^-$. For example $\sigma = \tilde \sigma$ yields the radar length deermined by the same polarization used for $N^+$ and $N^+$, while $\sigma =-\tilde \sigma$ represents the case when the reflection at the end of the object changes the polarization of the light.

\section{Relativistic Observables}
Having clarified an observers measurement of time and spatial lengths we can derive the classical relativistic observables time dilation and length contraction. To do so, consider two observers on inertial wordlines $x_1$ and $x_2$, which have met at some point in spacetime. There they synchronized their clocks. After a proper time $t_1$ the first observer may decompose the tangent of the worldline of the second observer in terms of an equal time displacement vector $X = t_1 V$ as $t_2 \dot x_2 = t_1 (\dot x_1 + V)$, where $V$ is the relative velocity between the observers. 

The relation between the proper time $t_2$ of the second observer and the proper time $t_1$ then is
\begin{align}
	t_2 L^*(\dot x_2)  = t_2= t_1 L^*(\dot x_1 + V) = t_1\sqrt{1 + \eta(V,V)}  \bigg(1 + \epsilon \frac{1}{12}\frac{h_1(V,V)}{(1 + \eta(V,V))}\bigg)\,.
\end{align}
Most interestingly, for weak premetric theories of electrodynamics for which $h_1^{ab} = \mathcal{K}^{acb}{}_d = 0$, no change in the time dilation between observers compared to special relativity appears.

The calculation of the length contraction is more involved. Let observer $x_1$ carry a spatial object $Y$ to which it associates the radar length $\mathcal{L}_{\dot x_1}(Y)$. With respect to the observer $x_2$ the object $Y$ is not spatial, so first one needs to determine the projection $\bar Y$ of $Y$ onto the spatial directions of $x_2$. Then, $x_2$ can associate the radar length $\mathcal{L}_{\dot x_2}(\bar Y)$ to $\bar Y$. The length contraction factor is given by the fraction $\frac{\mathcal{L}_{\dot x_1}(Y)}{\mathcal{L}_{\dot x_2}(\bar Y)}$. The result is a complicated function of the perturbations $h_1$ and $h_2$ which can be found in \cite{Gurlebeck:2018nme}. The interesting observation is that, in contracts to the time dilation, even for $h_1^{ab} = \mathcal{K}^{acb}{}_d = 0$ the effect does acquire a first order modification compared to the special relativistic result.

\section{Conclusion}
A modification of the geometry away from metric spacetime geometry requires a careful revision of the observer model used to compare observations with theoretical predictions. The example of the radar length demonstrates nicely that, in particular for Finslerian spacetime geometries it is not viable to simply exchange the Lorentzian metric employed in general relativity with the Finslerian metric evaluated at the observer curve on the tangent bundle for a description of observables.

The future task for spacetime geometries based on a Finsler function or a general dispersion relation is to find a description of observers from these quantities and to clarify the transformations mapping them onto each other.

\section*{Acknowledgments}
This talk is based on the article \cite{Gurlebeck:2018nme} written in collaboration with Norman G\"urlebeck.

\end{document}